\documentclass[aps,twocolumn,amsmath,amssymb,floatfixng,superscriptaddress,longbibliography,numerical]{revtex4-1}

\pdfoutput=1
\usepackage[dvips]{graphics}
\usepackage{bm}
\usepackage{epsfig}
\usepackage{enumerate}
\usepackage{subfigure}
\usepackage{color}
\usepackage{graphicx}
\usepackage{hyperref}
\hypersetup{
    pdfnewwindow=true,      
    colorlinks=true,       
    linkcolor=magenta,          
    citecolor=blue,        
    filecolor=blue,      
    urlcolor=blue        
}

\newcommand{\bea}{\begin{eqnarray}}
\newcommand{\eea}{\end{eqnarray}}
\newcommand{\beq}{\begin{equation}}  
\newcommand{\eeq}{\end{equation}}

\begin{document} 
\title{Photoinduced interfacial chiral modes in threefold topological semimetal} 

\author{SK Firoz Islam}
\email{firoz.seikh@aalto.fi}
\affiliation{Department of Applied Physics, Aalto University, P.~O.~Box 15100, FI-00076 AALTO, Finland}

\author{A.~A.~Zyuzin}
\affiliation{Department of Applied Physics, Aalto University, P.~O.~Box 15100, FI-00076 AALTO, Finland}
\affiliation{Ioffe Physical--Technical Institute,~194021 St.~Petersburg, Russia}

\begin{abstract}
We investigate the chiral electronic modes at the interface between two regions of a threefold topological semimetal, which is
illuminated by left and right handed elliptically polarized waves. The radiation effects on the band structure of semimetal is
analyzed by using Floquet theory. Two distinct solutions of the interface modes are found with the chirality depending on the phase of the irradiation.
We also consider the anomalous Hall response which is attributed to the transition between dispersionless flat band and conic bands.
\end{abstract}
\maketitle
\section{Introduction}
Recently, the possibility of manipulating the electronic band structure by applying a time dependent periodic perturbation
in the form of irradiation/light has received much attention, especially after the proposals of light induced topological phase
transition \cite{PhysRevB.79.081406,lindner2011floquet,PhysRevB.85.125425,PhysRevB.84.235108,PhysRevB.85.205428,PhysRevB.89.235416,PhysRevB.94.081103,
PhysRevLett.110.026603,PhysRevB.89.121401,cayssol2013floquet} (Floquet topological insulator) which has also been confirmed by experiments
\cite{peng2016experimental,zhang2014anomalous,wang2013observation}.
Apart from this, optical pumping can also be used to control the spin and the valley degree of freedom in Dirac materials \cite{PhysRevLett.116.016802,PhysRevB.84.195408,
PhysRevB.98.075422,PhysRevB.90.125438} which is the key requirement for the spin and valleytronics. Moreover, electromagnetic field can play the key role
in $0-\pi$ phase transition in Josephson current in a NSN hybrid junction made of silicene \cite{PhysRevB.94.165436}
and Weyl semimetal \cite{PhysRevB.95.201115}.

Topological semimetals with multiple bands have recently been predicted \cite{Bradlynaaf5037,PhysRevLett.119.206402}.
These materials are also known as multifold fermions and can be described by the higher number (more than two) of band crossing degeneracy,
which differs itself from the linear dispersion at the nodal band-touching points in Weyl semimetal. 
Several experiments \cite{PhysRevLett.122.076402,rao2019observation,
sanchez2019topological,schroter2019chiral} have also confirmed the existence of such multifold semimetals.
It has recently been found that multifold fermions are very promising in exhibiting the quantized photogalvanic
effect \cite{PhysRevLett.119.206401,PhysRevB.98.155145}. The signature of multifold bands in such materials has also been explored
in optical conductivity \cite{PhysRevB.99.155145}.

We particularly focus on three fold semimetal which can be considered as $3$D analog of $2$D dice lattice \cite{PhysRevB.84.195422,PhysRevLett.112.026402}.
One of the distinct features of three fold semimetal is the existence of dispersionless flat band which has some unusual consequences
in transport signature in its $2$D counterpart \cite{PhysRevB.88.161413,PhysRevB.92.035118,Biswas_2016}. The effects of light/irradiation
in $2$D dice lattice \cite{PhysRevB.98.075422,PhysRevB.99.205429} have been investigated very recently. On the other hand, much less attention has been paid for $3$D three
fold semimetal in the context of its interaction with irradiation. It can be easily anticipated that shedding light can induce a mass
term and subsequently opens a gap in the band dispersion, irrespective of the dimensionality. However, apart from the
gap opening, irradiation can also induce a momentum shift along the extra dimension and that shifting can 
be manipulated in opposite direction in two different regions of the materials by controlling the phase of the irradiation.
In this work, we follow this route to induce opposite momentum shifting in two different region by shedding irradiation
with opposite phase. The effects of irradiation is included by means of Floquet theory \cite{RevModPhys.89.011004} under
the limit of high frequency expansion (Floquet - Magnus expansion). We find that there exist two chiral interfacial surface modes and calculate the anomalous 
Hall conductivity induced due to irradiation.

The occurrence of analogues optical interfacial surface modes is well studied topic \cite{Yariv}. 
In optics, the interface of two optically active isotropic medias 
with opposite sign of the gyrotropic coefficient, supports unidirectional surface electromagnetic waves \cite{Seshadri, PhysRevB.61.12842, PhysRevB.92.115310}. 

The remainder of the paper is presented as follows. The section \ref{sec2} discusses the linearized Hamiltonian of the three band model and its band dispersion
with the effect of irradiation. The dispersion relations of chiral modes localized at the interface where the irradiation changes phase are obtained in section \ref{sec4}.
The section \ref{sec5} is devoted to calculation of the anomalous Hall conductivity and the transparency region for irradiation. Finally, we summarize and conclude in section \ref{sec6}.

\section{The model and Floquet theory}\label{sec2}
We start with the Hamiltonian, that describes the groups of SGs 199 and 214 materials as \cite{Bradlynaaf5037,PhysRevB.99.155145}
\begin{equation}
 H_{3f}=v\left[\begin{array}{ccc}
        0& e^{i\phi}k_z&e^{-i\phi}k_y\\
        e^{-i\phi}k_z & 0& e^{i\phi}k_x\\
        e^{i\phi}k_y &e^{-i\phi}k_x&0
       \end{array}\right],
\end{equation}
 where $\phi$ is a real parameter, $v$ is the Fermi velocity, and we use $c=k_B=\hbar=1$ throughout the calculations. We consider only 
 a single nodal point in the first Brillouin zone. The band is non-degenerate
 except at $k=0$, unless $\phi=n\pi/3$ with $n$ being integer.
 Another noticeable point is that for $\pi/3<\phi<2\pi/3$, the Hamiltonian is adiabatically connected to the one with $\phi=\pi/2$. 
 The Hamiltonian at this point can be written as $H_{3f}=v {\bf k.S}$ with $S_{x,y,z}$ being the generators of the rotation
 group $SO(2)$ in the pseudospin-1 representation defined as
  \begin{equation}
     \mathbf{S} = i\left[\begin{array}{ccc}
        0& \hat{e}_z& -\hat{e}_y\\
        -\hat{e}_z& 0& \hat{e}_x\\
        \hat{e}_y& -\hat{e}_x& 0
       \end{array}\right].
 \end{equation}
 
\begin{figure}[t]
\begin{minipage}{0.5\textwidth}
\includegraphics[width=.4\textwidth, height=4cm]{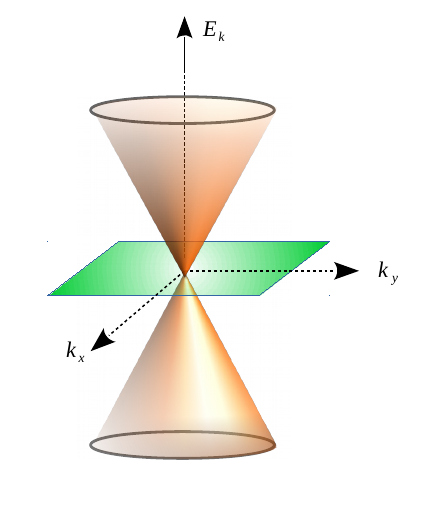}
\includegraphics[height=4cm,width=0.5\linewidth]{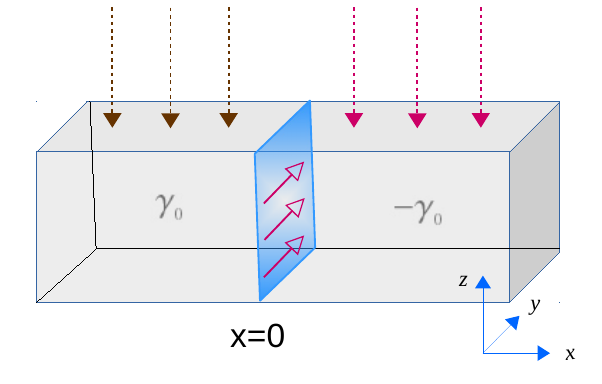}
\end{minipage}
\caption{(Left) A sketch of the energy dispersion for threefold semimetals for $k_z=0$. (Right) A schematic sketch of the setup. Two regions are illuminated by the elliptically polarized irradiation of opposite handedness.
Irradiation results in the domain wall at which parameter $\gamma = -\gamma_0[2\Theta(x)-1]$ changes sign. The interface where $\gamma=0$ is shown by blue 
plane at $x=0$. The domain wall hosts unidirectional electronic modes, 
with the chirality indicated by arrows in this plane.}
\label{band}
\end{figure}

The energy spectrum is given by dispersive bands $E_{\pm}=\pm v k$ and a disperionless flat band $E_0=0$, which is shown in
 Fig.~(\ref{band})a. This is a 3D analog of a 2D dice lattice.

Let us now consider that the system is subjected to an external time-dependent periodic perturbation,
which is described by a vector potential $\mathbf{A}(t)=[A_x\sin(\Omega t), A_y\sin(\Omega t-\delta), 0]$. Here $\Omega$ is the frequency of the 
irradiation and $\delta$ is the phase. The Hamiltonian describing threefold semimetal in the presence of radiation can be written by utilizing
the Peierls substitution $ \mathbf{k}\rightarrow \mathbf{k}-e\mathbf{A}(t)$, where $e<0$ is the electron charge, as
\begin{equation}
 H_{3f}(t)=H_{3f}+V(t),
\end{equation}
with
\begin{equation}
 V(t) =- ev[S_{x}A_x\sin(\Omega t) + S_yA_y\sin(\Omega t-\delta)].
\end{equation}
The above Hamiltonian can be solved by using Floquet theory which  states that the system under the time dependent periodic perturbation
exhibits a complete set of solutions of the form $\psi(r,t)=\phi(r,t)e^{-i\epsilon t}$ where $\phi(r,t)=\phi(r,t+T)$
and $T$ is the periodicity of the field,
the corresponding Floquet states  \cite{RevModPhys.89.011004}. It resembles the Bloch theorem in the momentum space. 
The Floquet eigenstates can be inserted into the time dependent Schrodinger equation to obtain 
the Floquet eigenvalue equation as $H_{3f}^{F}\phi(r,t)=\epsilon\phi(r,t)$
with $H^{F}_{3f} = H_{3f}(t)-i \partial_t$.

The Floquet eigenstates can be further expressed in Fourier form as
$\phi(r,t)=\sum_{n}\phi{_n}(r,t)e^{in\Omega t}$, where $n$ is the Fourier components  or Floquet side bands. To obtain the Floquet energy 
spectrum, the Floquet Hamiltonian has to be diagonalized in the basis of Floquet side band `$n$'. However, in such case one must restrict
the side band index by setting a cut-off. On the other hand, for high-frequency regime of the field, an effective Hamiltonian can be obtained
by following Floquet - Magnus expansion up to the second
order in field amplitude as \cite{RevModPhys.89.011004} 
\begin{equation}
 \tilde{H}_{3f}\simeq H_{3f}+\frac{[V_{+},V_{-}]}{\Omega},
\end{equation}
where the second term describes the irradiation induced correction with
\begin{equation}
 V_{m}=\frac{1}{T}\int_0^{T}V(t)e^{-im\Omega t}.
\end{equation}
The effective Hamiltonian can be further simplified to
\begin{equation}\label{gap}
 \tilde{H}_{3f}=H_{3f} + S_zv\gamma,
\end{equation}
where $\gamma=ve^2(2\Omega)^{-1} A_xA_y\sin\delta$. For linear polarizations, $\gamma=0$.
Note that the extra term in the Eq.~(\ref{gap}), $S_z\gamma$, causes a momentum shift along the $k_z$-direction
and that can be further seen in the energy spectrum of the irradiated system as $E_{\pm} = \pm v\sqrt{(k_z + \gamma)^2+k_x^2+k_y^2}$.

\section{Interfacial modes}\label{sec4}
\begin{figure}[t]
\centering
\includegraphics[height=6cm,width=\linewidth]{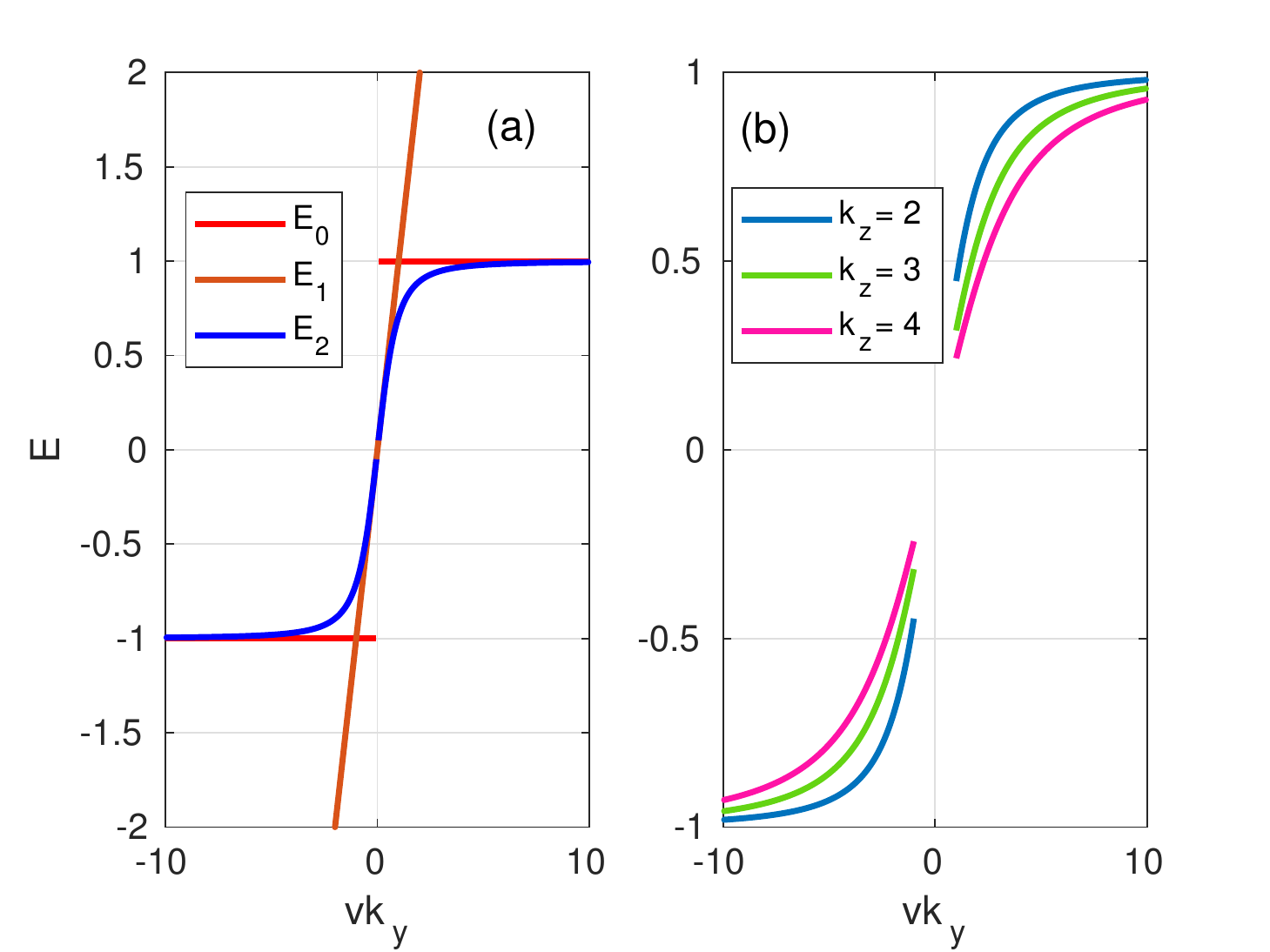}
\caption{At fixed $\gamma_0=1$ in the model of abrupt interface, the chiral interface modes given by the solution of Eq. \ref{sch} for 
two different momenta (a) $k_z\le 1$, where the dispersion of the chiral mode Eq. \ref{Chiral} stays intact $E_1=vk_y$, while the flat band Eq. \ref{FB} given by $E_0=v\gamma_0\mathrm{sgn}k_y$ at $k_z=0$ acquires the curvature $E_2$, and (b) $k_z>1$, where the mode Eq. \ref{Chiral} disappears.
}
\label{dispersion}
\end{figure}

In this section, we obtain the chiral modes propagating along the interface between two regions with different $\gamma$. 
The two regions of the material exposed to the external irradiation as schematically shown in Fig.~\ref{band}b.
The phase factor of the wave is adjusted in such way that $\gamma$ changes sign in two region as
\begin{equation}\label{gamma_def}
\gamma = -\gamma_0[2\Theta(x)-1],
\end{equation}
where $\gamma_0= ve^2 A_x A_y/2\Omega$, $\Theta(x)$ is the step function, and
$
\partial  \gamma/\partial x=-2\gamma_0\delta(x)
$. We seek for a solution of equation 
$\tilde{H}_{3f}\Psi = E\Psi$ in the form $\Psi=\left[\psi_1,\psi_2,\psi_3\right]^{T} e^{i(k_y y+k_z z)} $, which explicitly gives
\begin{equation}
 iv\left[\begin{array}{ccc}
        0& k_z+\gamma& - k_y\\
        -k_z-\gamma& 0& -i\partial_x\\
        k_y &i\partial_x&0
       \end{array}\right]\Psi = E\Psi.
\end{equation}
It is instructive to rewrite the above equation as
\begin{equation}\label{sch}
\frac{\partial^2 \psi_2}{\partial x^2}+U(x)\psi_2=\left\{ k_y^2+[k_z+\gamma(x)]^2- \frac{E^2}{v^2} \right\} \psi_2,
\end{equation}
where $U(x)=-2\gamma_0 vk_y\delta(x)/E$. The above equation resembles $1$D 
Schrodinger equation in a delta potential, which supports a bound state provided $\gamma_0 vk_y/E>0$.
By utilizing the continuity conditions for the wave-function at the interface $x=0$,
one obtains an equation to determine the dispersion relation
\begin{equation}\label{solkz}
\sum_{s=\pm} \left[k_y^2+(k_z+s\gamma_0)^2- \frac{E^2}{v^2}\right]^{1/2} = 2\gamma_0\frac{vk_y}{E}.
\end{equation}
Two solutions of this equation exist provided inequality $vk_y\gamma_0/E \geq 0$ holds:
\begin{subequations}
\begin{align}\label{Chiral}
&E_1 =  vk_y  \textrm{sgn}\gamma_0,\\\label{FB}
&E_2 =  \frac{v k_y \gamma_0}{\sqrt{k_y^2+k_z^2}}.
\end{align}
\end{subequations}
The first``Fermi-arc" like solution exists provided $|\gamma_0| \geq |k_z|$, while the second is determined in the region $k_y^2+k_z^2> |\gamma_0 k_z|$.
Note that the chirality of the modes is defined by the sign of parameter $\gamma_0$. The first dispersion is flat at $k_y=0$, while the second one becomes flat at $k_z=0$, as shown in Fig.~(\ref{dispersion})a. 
At finite values of momenta $k_z$ the second mode acquires a curvature with $k_y$
and this curvature increases further with $k_z$ as shown in Fig.~(\ref{dispersion})b. For completeness we present solution of Eq.~(\ref{solkz})
for finite $k_z$ in Fig.~(\ref{dispersion}). This figure contains
two subplots in which Fig.~(\ref{dispersion})b is for $k_z>1$ and Fig.~(\ref{dispersion})a is for $k_z\le 1$. 

The wave-function, which corresponds to the eigenvalue $E_{1,2}$, is proportional to $\Psi_{1,2} \propto e^{-x/\ell_{1,2;+}}$ at $x>0$ and
$\Psi_{1,2} \propto e^{x/\ell_{1,2;-}}$ at $x<0$. The localization lengths are given by $\ell_{1,\pm} = |k_z \mp \gamma_0|^{-1}$ 
and $\ell_{2,\pm} = |k_y^2+k_z^2 \mp \gamma_0 k_z|^{-1} \sqrt{k_y^2+k_z^2}$.

So far we have addressed the scenario for an abrupt interface between two regions. However, in realistic 
case, interface may not be always abrupt rather a smooth, which may be modelled as
\begin{equation}
 \gamma(x)=-\gamma_0\tanh(x/L), 
\end{equation} 
 where $L$ is the typical barrier width. This yields for $k_z=0$ the $1$D Schrodinger equation $
\partial^2_x \psi_2-U_0 \mathrm{sech}^2(x/L)\psi_2=( k_y^2+\gamma_0^2- E^2/v^2 ) \psi_2$, with
$
 U_0 = \frac{vk_y}{E}\frac{\gamma_0}{L} - \gamma_0^2.
$
By following the standard solution \cite{Landau},
\begin{eqnarray}
\frac{E^2}{v^2} -k_y^2-\gamma_0^2=\frac{L^2}{4}\left[\sqrt{1+4L^2U_0}-(1+2n)\right]^2,~~~
\end{eqnarray}
where $n=0,1,2,3...$ The dispersion of the interface modes can be found by solving the above equation, which
hints that there exist many solutions depending on the values of $n< (-1+\sqrt{1+ 4L^2 U_0})/2$ for $k_z=0$.

The basic recipe of engineering such interfacial modes is the momentum shift along the $k_z$ direction by applying irradiation.

It is worthwhile to mention at this stage that topological $1$D chiral mode was predicted on the
$2$D surface of a $3$D topological insulator (TI) \cite{PhysRevB.91.241404}. The irradiation induced band gap 
opening on the surface of $3$D TI plays the key role there. The present study is based on $3$D
materials, where the external irradiation results in the momentum shift along the extra dimension
and formation of the Fermi arc interface modes. Additionally, a finite number of propagating $2$D
interface modes evolved depending on the interface width, which can be found by simple quantum mechanical treatment.

\section{Anomalous Hall response}\label{sec5}
Let us now briefly discuss signatures of the irradiation in the response function. We particularly focus on anomalous Hall conductivity,
which is exclusively attributed to the irradiation unlike longitudinal conductivity. 

We should mention here that the expression for the current-current operator in the presence of the intense AC background has been formulated with the Floquet method
in \cite{PhysRevB.79.081406,PhysRevB.98.205109}. It was shown that the time averaged operator can be written in terms of the Green function, where the energies are replaced with the Floquet quasi-energies
\begin{equation}\label{anoma}
\Pi_{xy}(\omega_m)= \textrm{tr}\sum_{n}T\int \frac{d^3k}{(2\pi)^3}G(i\omega_n+i\omega_m,\mathbf{k})
j_{x} G(i\omega_n,\mathbf{k})j_{y},~~~~~
\end{equation} 
where the Green function is given by 
\begin{eqnarray}\label{green}
 G(i\omega_n,\mathbf{k}) &=& [i\omega_n+\mu-v{\bf S.k}]^{-1}
 \\\nonumber
 &=&
 \frac{M_0(\mathbf{k})}{i\omega_n+\mu}+\frac{1}{2}\sum_{s=\pm}\frac{1-M_{0}(\mathbf{k}) - isM_{1}(\mathbf{k})}{i\omega_n+\mu-svk},~~~
\end{eqnarray}
with $\mu$ is the chemical potential, and $\omega_n, \omega_m$ are the fermionic and bosonic Matsubara frequencies with $n, m \in \mathcal{Z}$. 
The different components of Green function are given by
\begin{equation}
 M_{0}(\mathbf{k})=\frac{1}{k^2}\left[\begin{array}{ccc}
        k_x^2 &k_xk_y &k_xk_z\\
        k_xk_y & k_y^2& k_yk_z\\
        k_xk_z &k_yk_z&k_z^2
       \end{array}\right],
\end{equation}
and
\begin{equation}
 M_{1}(\mathbf{k})=\frac{1}{k}\left[\begin{array}{ccc}
        0 &k_z &k_y\\
        k_z & 0& k_x\\
        k_y &k_x&0
       \end{array}\right].
\end{equation}
The first term in Eq.~(\ref{green}) corresponds to the dispersionless flat band whereas the second term corresponds
to the conic bands. The components of the current density operator are given by
$
j_{\alpha} = e\partial \tilde{H}_{3f}/\partial k_{\alpha}.
$
At zero temperature, using $k=\sqrt{k_{\bot}^2+k_z^2}$,
$
\int\frac{d^3k}{(2\pi)^3}...=  \int_{-\Lambda_{\textrm{reg}}-\gamma_0}^{\Lambda_\textrm{reg}+\gamma_0}\frac{dk_z}{2\pi}\int
\frac{d^2k_{\bot}}{4\pi^2}...
$ with $\Lambda_{\textrm{reg}}$ introduced for the correct definition of the $k_z$ integral, and performing analytical continuation $i\omega_m \rightarrow \omega + i\delta$, it is straight forward to arrive at
the anomalous Hall conductivity $\sigma_{xy} = \lim_{\omega \to 0} \Pi_{xy}(\omega)/i\omega$ as
\begin{equation}
 \sigma_{xy} = \frac{e^2 \gamma_0}{2\pi^2},
\end{equation}
which is two times larger than anomalous Hall conductivity in the Weyl semimetal due to the doubling of the topological charge.
The sign of $\sigma_{xy}$ depends on the phase of the incident radiation. Note that it is the transition between the dispersionless flat band and conic band which causes such result. The inter or intra conic band
transition does not contribute to the anomalous Hall conductivity, as the corresponding matrix element vanishes in the polarization operator in Eq.~(\ref{anoma}).
We also mention here that the above formalism is well justified as long as we treat the irradiated Hamiltonian via Floquet-Magnus high frequency approximation,
where only the two nearest side bands are taken into account. It is instructive to adjust the chemical potential near to the band touching point in order to avoid
any interference of higher Floquet side bands. To capture higher side bands, one must consider beyond high frequency approximation and
Floquet Green function approach for transport study \cite{PhysRevB.79.081406}.

For completeness, we also comment on the diagonal components of the polarization matrix at zero temperature and wave-vector, which reduces to
\begin{eqnarray}
  \Pi_{xx} =  \frac{ \omega_{m}^2 }{6\pi^2 v} \Big[\ln\frac{\Lambda^2}{\mu^2+\omega_m^2}+\frac{\mu^2}{\omega_m^2}\Big],
\end{eqnarray}
where $\Lambda$ is the energy cut-off and the diamagnetic contribution has been subtracted. 
The effect of flat band is to render the valence to conduction inter-band transitions, which leads to imaginary component
already at $\omega = \mu$. This is in contrast to $\omega = 2\mu$ condition in 3D Weyl semimetals.
The transparency region to the external irradiation can be estimated by considering the conditions for the zero imaginary part and positive real components of the dielectric function. These
require frequencies of the incident wave to be confined in the interval
\begin{equation}
\mu> \omega\ge\frac{\alpha \mu}{(1+\alpha^2\ln|\Lambda/\mu\sqrt{1-\alpha^2}| )^{1/2}},
\end{equation} 
with $\alpha=N/(6\pi^2v\epsilon_0)$ in which $\epsilon_0$ is the permittivity of free space and $N$ is the number of threefold band touching points.

\section{Discussion and conclusions}\label{sec6}
Let us estimate the parameters of the model.
Typically, one can use the standard parameters of irradiation in the high frequency limit as: $evA=0.1-1$ eV and 
$0.1<evA/\Omega<1$.
Under this regime, we can estimate 
$v\gamma_0= v^2e^2 A_x A_y/2\Omega \approx 0.5(eAv/2)=0.12$ eV.
The frequency $\Omega$ can be estimated to be $\approx 200$ THz for $eAv=0.5$ eV, which gives the wavelength of the irradiation as $1.2$ $\mu$m.

To summarize, we present a theoretical proposal of engineering a unidirectional mode propagating along the interface between two regions of threefold
semimetal. The band structure of such material in the presence of the irradiation is obtained within the Floquet theory in the high frequency limit 
treating external field as a perturbation within the Floquet - Magnus expansion. 

It is shown that time-dependent periodic perturbation, in the form of 
elliptically polarized field in $x-y$ plane, can cause a momentum shift along the $k_z$ direction in the electronic band structure. 
By utilizing several sources of irradiation, domain walls might be realized in the material at which phase factor of field changes. 
It is found that the interface between two regions, exposed to the irradiation with opposite phase, 
can host unidirectional Fermi-arc mode in addition to a dispersionless flat mode. 
The anomalous Hall conductivity and frequency interval at which the material becomes transparent to irradiation by evaluating the dielectric tensor are discussed.

{\bf Acknowledgements}
This work is supported by the Academy of Finland. A.A.Z. is grateful to the hospitality of the Pirinem School of Theoretical Physics. 

\bibliography{bibfile_threefold}

\begin{thebibliography}{42}%
\makeatletter
\providecommand \@ifxundefined [1]{%
 \@ifx{#1\undefined}
}%
\providecommand \@ifnum [1]{%
 \ifnum #1\expandafter \@firstoftwo
 \else \expandafter \@secondoftwo
 \fi
}%
\providecommand \@ifx [1]{%
 \ifx #1\expandafter \@firstoftwo
 \else \expandafter \@secondoftwo
 \fi
}%
\providecommand \natexlab [1]{#1}%
\providecommand \enquote  [1]{``#1''}%
\providecommand \bibnamefont  [1]{#1}%
\providecommand \bibfnamefont [1]{#1}%
\providecommand \citenamefont [1]{#1}%
\providecommand \href@noop [0]{\@secondoftwo}%
\providecommand \href [0]{\begingroup \@sanitize@url \@href}%
\providecommand \@href[1]{\@@startlink{#1}\@@href}%
\providecommand \@@href[1]{\endgroup#1\@@endlink}%
\providecommand \@sanitize@url [0]{\catcode `\\12\catcode `\$12\catcode
  `\&12\catcode `\#12\catcode `\^12\catcode `\_12\catcode `\%12\relax}%
\providecommand \@@startlink[1]{}%
\providecommand \@@endlink[0]{}%
\providecommand \url  [0]{\begingroup\@sanitize@url \@url }%
\providecommand \@url [1]{\endgroup\@href {#1}{\urlprefix }}%
\providecommand \urlprefix  [0]{URL }%
\providecommand \Eprint [0]{\href }%
\providecommand \doibase [0]{http://dx.doi.org/}%
\providecommand \selectlanguage [0]{\@gobble}%
\providecommand \bibinfo  [0]{\@secondoftwo}%
\providecommand \bibfield  [0]{\@secondoftwo}%
\providecommand \translation [1]{[#1]}%
\providecommand \BibitemOpen [0]{}%
\providecommand \bibitemStop [0]{}%
\providecommand \bibitemNoStop [0]{.\EOS\space}%
\providecommand \EOS [0]{\spacefactor3000\relax}%
\providecommand \BibitemShut  [1]{\csname bibitem#1\endcsname}%
\let\auto@bib@innerbib\@empty
\bibitem [{\citenamefont {Oka}\ and\ \citenamefont
  {Aok}(2009)}]{PhysRevB.79.081406}%
  \BibitemOpen
  \bibfield  {author} {\bibinfo {author} {\bibfnamefont {T.}~\bibnamefont
  {Oka}}\ and\ \bibinfo {author} {\bibfnamefont {H.i}\ \bibnamefont {Aok}},\
  }\bibfield  {title} {\enquote {\bibinfo {title} {{Photovoltaic Hall effect in
  graphene}},}\ }\href {\doibase 10.1103/PhysRevB.79.081406} {\bibfield
  {journal} {\bibinfo  {journal} {Phys. Rev. B}\ }\textbf {\bibinfo {volume}
  {79}},\ \bibinfo {pages} {081406} (\bibinfo {year} {2009})}\BibitemShut
  {NoStop}%
\bibitem [{\citenamefont {Lindner}\ \emph {et~al.}(2011)\citenamefont
  {Lindner}, \citenamefont {Refael},\ and\ \citenamefont
  {Galitski}}]{lindner2011floquet}%
  \BibitemOpen
  \bibfield  {author} {\bibinfo {author} {\bibfnamefont {N~H}\ \bibnamefont
  {Lindner}}, \bibinfo {author} {\bibfnamefont {G.}~\bibnamefont {Refael}}, \
  and\ \bibinfo {author} {\bibfnamefont {V.}~\bibnamefont {Galitski}},\
  }\bibfield  {title} {\enquote {\bibinfo {title} {{Floquet topological
  insulator in semiconductor quantum wells}},}\ }\href {\doibase
  10.1038/nphys1926} {\bibfield  {journal} {\bibinfo  {journal} {Nature Phys.}\
  }\textbf {\bibinfo {volume} {7}},\ \bibinfo {pages} {490} (\bibinfo {year}
  {2011})}\BibitemShut {NoStop}%
\bibitem [{\citenamefont {Inoue}\ and\ \citenamefont
  {Tanaka}(2012)}]{PhysRevB.85.125425}%
  \BibitemOpen
  \bibfield  {author} {\bibinfo {author} {\bibfnamefont {J.}~\bibnamefont
  {Inoue}}\ and\ \bibinfo {author} {\bibfnamefont {A.}~\bibnamefont {Tanaka}},\
  }\bibfield  {title} {\enquote {\bibinfo {title} {{Photoinduced spin Chern
  number change in a two-dimensional quantum spin Hall insulator with broken
  spin rotational symmetry}},}\ }\href {\doibase 10.1103/PhysRevB.85.125425}
  {\bibfield  {journal} {\bibinfo  {journal} {Phys. Rev. B}\ }\textbf {\bibinfo
  {volume} {85}},\ \bibinfo {pages} {125425} (\bibinfo {year}
  {2012})}\BibitemShut {NoStop}%
\bibitem [{\citenamefont {Kitagawa}\ \emph {et~al.}(2011)\citenamefont
  {Kitagawa}, \citenamefont {Oka}, \citenamefont {Brataas}, \citenamefont
  {Fu},\ and\ \citenamefont {Demler}}]{PhysRevB.84.235108}%
  \BibitemOpen
  \bibfield  {author} {\bibinfo {author} {\bibfnamefont {T.}~\bibnamefont
  {Kitagawa}}, \bibinfo {author} {\bibfnamefont {T.}~\bibnamefont {Oka}},
  \bibinfo {author} {\bibfnamefont {A.}~\bibnamefont {Brataas}}, \bibinfo
  {author} {\bibfnamefont {L.}~\bibnamefont {Fu}}, \ and\ \bibinfo {author}
  {\bibfnamefont {E.}~\bibnamefont {Demler}},\ }\bibfield  {title} {\enquote
  {\bibinfo {title} {{Transport properties of nonequilibrium systems under the
  application of light: Photoinduced quantum Hall insulators without Landau
  levels}},}\ }\href {\doibase 10.1103/PhysRevB.84.235108} {\bibfield
  {journal} {\bibinfo  {journal} {Phys. Rev. B}\ }\textbf {\bibinfo {volume}
  {84}},\ \bibinfo {pages} {235108} (\bibinfo {year} {2011})}\BibitemShut
  {NoStop}%
\bibitem [{\citenamefont {L\'opez}\ \emph {et~al.}(2012)\citenamefont
  {L\'opez}, \citenamefont {Sun},\ and\ \citenamefont
  {Schliemann}}]{PhysRevB.85.205428}%
  \BibitemOpen
  \bibfield  {author} {\bibinfo {author} {\bibfnamefont {A.}~\bibnamefont
  {L\'opez}}, \bibinfo {author} {\bibfnamefont {Z.~Z.}\ \bibnamefont {Sun}}, \
  and\ \bibinfo {author} {\bibfnamefont {J.}~\bibnamefont {Schliemann}},\
  }\bibfield  {title} {\enquote {\bibinfo {title} {{Floquet spin states in
  graphene under ac-driven spin-orbit interaction}},}\ }\href {\doibase
  10.1103/PhysRevB.85.205428} {\bibfield  {journal} {\bibinfo  {journal} {Phys.
  Rev. B}\ }\textbf {\bibinfo {volume} {85}},\ \bibinfo {pages} {205428}
  (\bibinfo {year} {2012})}\BibitemShut {NoStop}%
\bibitem [{\citenamefont {Zhai}\ and\ \citenamefont
  {Jin}(2014)}]{PhysRevB.89.235416}%
  \BibitemOpen
  \bibfield  {author} {\bibinfo {author} {\bibfnamefont {X.}~\bibnamefont
  {Zhai}}\ and\ \bibinfo {author} {\bibfnamefont {G.}~\bibnamefont {Jin}},\
  }\bibfield  {title} {\enquote {\bibinfo {title} {{Photoinduced topological
  phase transition in epitaxial graphene}},}\ }\href {\doibase
  10.1103/PhysRevB.89.235416} {\bibfield  {journal} {\bibinfo  {journal} {Phys.
  Rev. B}\ }\textbf {\bibinfo {volume} {89}},\ \bibinfo {pages} {235416}
  (\bibinfo {year} {2014})}\BibitemShut {NoStop}%
\bibitem [{\citenamefont {Saha}(2016)}]{PhysRevB.94.081103}%
  \BibitemOpen
  \bibfield  {author} {\bibinfo {author} {\bibfnamefont {K.}~\bibnamefont
  {Saha}},\ }\bibfield  {title} {\enquote {\bibinfo {title} {{Photoinduced
  Chern insulating states in semi-Dirac materials}},}\ }\href {\doibase
  10.1103/PhysRevB.94.081103} {\bibfield  {journal} {\bibinfo  {journal} {Phys.
  Rev. B}\ }\textbf {\bibinfo {volume} {94}},\ \bibinfo {pages} {081103}
  (\bibinfo {year} {2016})}\BibitemShut {NoStop}%
\bibitem [{\citenamefont {Ezawa}(2013)}]{PhysRevLett.110.026603}%
  \BibitemOpen
  \bibfield  {author} {\bibinfo {author} {\bibfnamefont {M.}~\bibnamefont
  {Ezawa}},\ }\bibfield  {title} {\enquote {\bibinfo {title} {{Photoinduced
  Topological Phase Transition and a Single Dirac-Cone State in Silicene}},}\
  }\href {\doibase 10.1103/PhysRevLett.110.026603} {\bibfield  {journal}
  {\bibinfo  {journal} {Phys. Rev. Lett.}\ }\textbf {\bibinfo {volume} {110}},\
  \bibinfo {pages} {026603} (\bibinfo {year} {2013})}\BibitemShut {NoStop}%
\bibitem [{\citenamefont {Perez-Piskunow}\ \emph {et~al.}(2014)\citenamefont
  {Perez-Piskunow}, \citenamefont {Usaj}, \citenamefont {Balseiro},\ and\
  \citenamefont {Torres}}]{PhysRevB.89.121401}%
  \BibitemOpen
  \bibfield  {author} {\bibinfo {author} {\bibfnamefont {P.~M.}\ \bibnamefont
  {Perez-Piskunow}}, \bibinfo {author} {\bibfnamefont {G.}~\bibnamefont
  {Usaj}}, \bibinfo {author} {\bibfnamefont {C.~A.}\ \bibnamefont {Balseiro}},
  \ and\ \bibinfo {author} {\bibfnamefont {L.~E. F.~F.}\ \bibnamefont
  {Torres}},\ }\bibfield  {title} {\enquote {\bibinfo {title} {{Floquet chiral
  edge states in graphene}},}\ }\href {\doibase 10.1103/PhysRevB.89.121401}
  {\bibfield  {journal} {\bibinfo  {journal} {Phys. Rev. B}\ }\textbf {\bibinfo
  {volume} {89}},\ \bibinfo {pages} {121401} (\bibinfo {year}
  {2014})}\BibitemShut {NoStop}%
\bibitem [{\citenamefont {Cayssol}\ \emph {et~al.}(2013)\citenamefont
  {Cayssol}, \citenamefont {D{\'o}ra}, \citenamefont {Simon},\ and\
  \citenamefont {Moessner}}]{cayssol2013floquet}%
  \BibitemOpen
  \bibfield  {author} {\bibinfo {author} {\bibfnamefont {J.}~\bibnamefont
  {Cayssol}}, \bibinfo {author} {\bibfnamefont {B.}~\bibnamefont {D{\'o}ra}},
  \bibinfo {author} {\bibfnamefont {F.}~\bibnamefont {Simon}}, \ and\ \bibinfo
  {author} {\bibfnamefont {R.}~\bibnamefont {Moessner}},\ }\bibfield  {title}
  {\enquote {\bibinfo {title} {{Floquet topological insulators}},}\ }\href
  {\doibase 10.1002/pssr.201206451} {\bibfield  {journal} {\bibinfo  {journal}
  {Phys. Stat. Sol. (RRL)}\ }\textbf {\bibinfo {volume} {7}},\ \bibinfo {pages}
  {101} (\bibinfo {year} {2013})}\BibitemShut {NoStop}%
\bibitem [{\citenamefont {Peng}\ \emph {et~al.}(2016)\citenamefont {Peng},
  \citenamefont {Qin}, \citenamefont {Zhao}, \citenamefont {Shen},
  \citenamefont {Xu}, \citenamefont {Bao}, \citenamefont {Jia},\ and\
  \citenamefont {Zhu}}]{peng2016experimental}%
  \BibitemOpen
  \bibfield  {author} {\bibinfo {author} {\bibfnamefont {Y.-G.}\ \bibnamefont
  {Peng}}, \bibinfo {author} {\bibfnamefont {C.-Z.}\ \bibnamefont {Qin}},
  \bibinfo {author} {\bibfnamefont {D.-G.}\ \bibnamefont {Zhao}}, \bibinfo
  {author} {\bibfnamefont {Y.-X.}\ \bibnamefont {Shen}}, \bibinfo {author}
  {\bibfnamefont {X.-Y.}\ \bibnamefont {Xu}}, \bibinfo {author} {\bibfnamefont
  {M.}~\bibnamefont {Bao}}, \bibinfo {author} {\bibfnamefont {H.}~\bibnamefont
  {Jia}}, \ and\ \bibinfo {author} {\bibfnamefont {X.-F.}\ \bibnamefont
  {Zhu}},\ }\bibfield  {title} {\enquote {\bibinfo {title} {{Experimental
  demonstration of anomalous Floquet topological insulator for sound}},}\
  }\href {\doibase 10.1038/ncomms13368} {\bibfield  {journal} {\bibinfo
  {journal} {Nature Com.}\ }\textbf {\bibinfo {volume} {7}},\ \bibinfo {pages}
  {13368} (\bibinfo {year} {2016})}\BibitemShut {NoStop}%
\bibitem [{\citenamefont {Zhang}\ \emph {et~al.}(2014)\citenamefont {Zhang},
  \citenamefont {Yao}, \citenamefont {Shao}, \citenamefont {Li}, \citenamefont
  {Li}, \citenamefont {Bao}, \citenamefont {Wang},\ and\ \citenamefont
  {Yang}}]{zhang2014anomalous}%
  \BibitemOpen
  \bibfield  {author} {\bibinfo {author} {\bibfnamefont {H.}~\bibnamefont
  {Zhang}}, \bibinfo {author} {\bibfnamefont {J.}~\bibnamefont {Yao}}, \bibinfo
  {author} {\bibfnamefont {J.}~\bibnamefont {Shao}}, \bibinfo {author}
  {\bibfnamefont {H.}~\bibnamefont {Li}}, \bibinfo {author} {\bibfnamefont
  {S.}~\bibnamefont {Li}}, \bibinfo {author} {\bibfnamefont {D.}~\bibnamefont
  {Bao}}, \bibinfo {author} {\bibfnamefont {C.}~\bibnamefont {Wang}}, \ and\
  \bibinfo {author} {\bibfnamefont {G.}~\bibnamefont {Yang}},\ }\bibfield
  {title} {\enquote {\bibinfo {title} {{Anomalous photoelectric effect of a
  polycrystalline topological insulator film}},}\ }\href {\doibase
  10.1038/srep05876} {\bibfield  {journal} {\bibinfo  {journal} {Sci. rep.}\
  }\textbf {\bibinfo {volume} {4}},\ \bibinfo {pages} {5876} (\bibinfo {year}
  {2014})}\BibitemShut {NoStop}%
\bibitem [{\citenamefont {Wang}\ \emph {et~al.}(2013)\citenamefont {Wang},
  \citenamefont {Steinberg}, \citenamefont {Jarillo-Herrero},\ and\
  \citenamefont {Gedik}}]{wang2013observation}%
  \BibitemOpen
  \bibfield  {author} {\bibinfo {author} {\bibfnamefont {Y.}~\bibnamefont
  {Wang}}, \bibinfo {author} {\bibfnamefont {H.}~\bibnamefont {Steinberg}},
  \bibinfo {author} {\bibfnamefont {P.}~\bibnamefont {Jarillo-Herrero}}, \ and\
  \bibinfo {author} {\bibfnamefont {N.}~\bibnamefont {Gedik}},\ }\bibfield
  {title} {\enquote {\bibinfo {title} {{Observation of Floquet-Bloch states on
  the surface of a topological insulator}},}\ }\href {\doibase
  10.1126/science.1239834} {\bibfield  {journal} {\bibinfo  {journal}
  {Science}\ }\textbf {\bibinfo {volume} {342}},\ \bibinfo {pages} {453--457}
  (\bibinfo {year} {2013})}\BibitemShut {NoStop}%
\bibitem [{\citenamefont {Kundu}\ \emph {et~al.}(2016)\citenamefont {Kundu},
  \citenamefont {Fertig},\ and\ \citenamefont
  {Seradjeh}}]{PhysRevLett.116.016802}%
  \BibitemOpen
  \bibfield  {author} {\bibinfo {author} {\bibfnamefont {A.}~\bibnamefont
  {Kundu}}, \bibinfo {author} {\bibfnamefont {H.~A.}\ \bibnamefont {Fertig}}, \
  and\ \bibinfo {author} {\bibfnamefont {B.}~\bibnamefont {Seradjeh}},\
  }\bibfield  {title} {\enquote {\bibinfo {title} {{Floquet-Engineered
  Valleytronics in Dirac Systems}},}\ }\href {\doibase
  10.1103/PhysRevLett.116.016802} {\bibfield  {journal} {\bibinfo  {journal}
  {Phys. Rev. Lett.}\ }\textbf {\bibinfo {volume} {116}},\ \bibinfo {pages}
  {016802} (\bibinfo {year} {2016})}\BibitemShut {NoStop}%
\bibitem [{\citenamefont {Golub}\ \emph {et~al.}(2011)\citenamefont {Golub},
  \citenamefont {Tarasenko}, \citenamefont {Entin},\ and\ \citenamefont
  {Magarill}}]{PhysRevB.84.195408}%
  \BibitemOpen
  \bibfield  {author} {\bibinfo {author} {\bibfnamefont {L.~E.}\ \bibnamefont
  {Golub}}, \bibinfo {author} {\bibfnamefont {S.~A.}\ \bibnamefont
  {Tarasenko}}, \bibinfo {author} {\bibfnamefont {M.~V.}\ \bibnamefont
  {Entin}}, \ and\ \bibinfo {author} {\bibfnamefont {L.~I.}\ \bibnamefont
  {Magarill}},\ }\bibfield  {title} {\enquote {\bibinfo {title} {{Valley
  separation in graphene by polarized light}},}\ }\href {\doibase
  10.1103/PhysRevB.84.195408} {\bibfield  {journal} {\bibinfo  {journal} {Phys.
  Rev. B}\ }\textbf {\bibinfo {volume} {84}},\ \bibinfo {pages} {195408}
  (\bibinfo {year} {2011})}\BibitemShut {NoStop}%
\bibitem [{\citenamefont {Dey}\ and\ \citenamefont
  {Ghosh}(2018)}]{PhysRevB.98.075422}%
  \BibitemOpen
  \bibfield  {author} {\bibinfo {author} {\bibfnamefont {B.}~\bibnamefont
  {Dey}}\ and\ \bibinfo {author} {\bibfnamefont {T.~K.}\ \bibnamefont
  {Ghosh}},\ }\bibfield  {title} {\enquote {\bibinfo {title} {{Photoinduced
  valley and electron-hole symmetry breaking in $\alpha-{T}_{3}$ lattice: The
  role of a variable Berry phase}},}\ }\href {\doibase
  10.1103/PhysRevB.98.075422} {\bibfield  {journal} {\bibinfo  {journal} {Phys.
  Rev. B}\ }\textbf {\bibinfo {volume} {98}},\ \bibinfo {pages} {075422}
  (\bibinfo {year} {2018})}\BibitemShut {NoStop}%
\bibitem [{\citenamefont {Tahir}\ \emph {et~al.}(2014)\citenamefont {Tahir},
  \citenamefont {Manchon},\ and\ \citenamefont
  {Schwingenschl\"ogl}}]{PhysRevB.90.125438}%
  \BibitemOpen
  \bibfield  {author} {\bibinfo {author} {\bibfnamefont {M.}~\bibnamefont
  {Tahir}}, \bibinfo {author} {\bibfnamefont {A.}~\bibnamefont {Manchon}}, \
  and\ \bibinfo {author} {\bibfnamefont {U.}~\bibnamefont
  {Schwingenschl\"ogl}},\ }\bibfield  {title} {\enquote {\bibinfo {title}
  {{Photoinduced quantum spin and valley Hall effects, and orbital
  magnetization in monolayer ${\mathrm{MoS}}_{2}$}},}\ }\href {\doibase
  10.1103/PhysRevB.90.125438} {\bibfield  {journal} {\bibinfo  {journal} {Phys.
  Rev. B}\ }\textbf {\bibinfo {volume} {90}},\ \bibinfo {pages} {125438}
  (\bibinfo {year} {2014})}\BibitemShut {NoStop}%
\bibitem [{\citenamefont {Zhou}\ and\ \citenamefont
  {Jin}(2016)}]{PhysRevB.94.165436}%
  \BibitemOpen
  \bibfield  {author} {\bibinfo {author} {\bibfnamefont {X.}~\bibnamefont
  {Zhou}}\ and\ \bibinfo {author} {\bibfnamefont {G.}~\bibnamefont {Jin}},\
  }\bibfield  {title} {\enquote {\bibinfo {title} {{Light-modulated
  0-$\ensuremath{\pi}$ transition in a silicene-based Josephson junction}},}\
  }\href {\doibase 10.1103/PhysRevB.94.165436} {\bibfield  {journal} {\bibinfo
  {journal} {Phys. Rev. B}\ }\textbf {\bibinfo {volume} {94}},\ \bibinfo
  {pages} {165436} (\bibinfo {year} {2016})}\BibitemShut {NoStop}%
\bibitem [{\citenamefont {Khanna}\ \emph {et~al.}(2017)\citenamefont {Khanna},
  \citenamefont {Rao},\ and\ \citenamefont {Kundu}}]{PhysRevB.95.201115}%
  \BibitemOpen
  \bibfield  {author} {\bibinfo {author} {\bibfnamefont {U.}~\bibnamefont
  {Khanna}}, \bibinfo {author} {\bibfnamefont {S.}~\bibnamefont {Rao}}, \ and\
  \bibinfo {author} {\bibfnamefont {A.}~\bibnamefont {Kundu}},\ }\bibfield
  {title} {\enquote {\bibinfo {title}
  {{$0\text{\ensuremath{-}}\ensuremath{\pi}$ transitions in a Josephson
  junction of an irradiated Weyl semimetal}},}\ }\href {\doibase
  10.1103/PhysRevB.95.201115} {\bibfield  {journal} {\bibinfo  {journal} {Phys.
  Rev. B}\ }\textbf {\bibinfo {volume} {95}},\ \bibinfo {pages} {201115}
  (\bibinfo {year} {2017})}\BibitemShut {NoStop}%
\bibitem [{\citenamefont {Bradlyn}\ \emph {et~al.}(2016)\citenamefont
  {Bradlyn}, \citenamefont {Cano}, \citenamefont {Wang}, \citenamefont
  {Vergniory}, \citenamefont {Felser}, \citenamefont {Cava},\ and\
  \citenamefont {Bernevig}}]{Bradlynaaf5037}%
  \BibitemOpen
  \bibfield  {author} {\bibinfo {author} {\bibfnamefont {B.}~\bibnamefont
  {Bradlyn}}, \bibinfo {author} {\bibfnamefont {J.}~\bibnamefont {Cano}},
  \bibinfo {author} {\bibfnamefont {Z.}~\bibnamefont {Wang}}, \bibinfo {author}
  {\bibfnamefont {M.~G.}\ \bibnamefont {Vergniory}}, \bibinfo {author}
  {\bibfnamefont {C.}~\bibnamefont {Felser}}, \bibinfo {author} {\bibfnamefont
  {R.~J.}\ \bibnamefont {Cava}}, \ and\ \bibinfo {author} {\bibfnamefont
  {B.~A.}\ \bibnamefont {Bernevig}},\ }\bibfield  {title} {\enquote {\bibinfo
  {title} {{Beyond Dirac and Weyl fermions: Unconventional quasiparticles in
  conventional crystals}},}\ }\href {\doibase 10.1126/science.aaf5037}
  {\bibfield  {journal} {\bibinfo  {journal} {Science}\ }\textbf {\bibinfo
  {volume} {353}},\ \bibinfo {pages} {6299} (\bibinfo {year}
  {2016})}\BibitemShut {NoStop}%
\bibitem [{\citenamefont {Tang}\ \emph {et~al.}(2017)\citenamefont {Tang},
  \citenamefont {Zhou},\ and\ \citenamefont {Zhang}}]{PhysRevLett.119.206402}%
  \BibitemOpen
  \bibfield  {author} {\bibinfo {author} {\bibfnamefont {P.}~\bibnamefont
  {Tang}}, \bibinfo {author} {\bibfnamefont {Q.}~\bibnamefont {Zhou}}, \ and\
  \bibinfo {author} {\bibfnamefont {S.-C.}\ \bibnamefont {Zhang}},\ }\bibfield
  {title} {\enquote {\bibinfo {title} {{Multiple Types of Topological Fermions
  in Transition Metal Silicides}},}\ }\href {\doibase
  10.1103/PhysRevLett.119.206402} {\bibfield  {journal} {\bibinfo  {journal}
  {Phys. Rev. Lett.}\ }\textbf {\bibinfo {volume} {119}},\ \bibinfo {pages}
  {206402} (\bibinfo {year} {2017})}\BibitemShut {NoStop}%
\bibitem [{\citenamefont {Takane}\ \emph {et~al.}(2019)\citenamefont {Takane},
  \citenamefont {Wang}, \citenamefont {Souma},\ and\ \citenamefont
  {et~al.}}]{PhysRevLett.122.076402}%
  \BibitemOpen
  \bibfield  {author} {\bibinfo {author} {\bibfnamefont {D.}~\bibnamefont
  {Takane}}, \bibinfo {author} {\bibfnamefont {Z.}~\bibnamefont {Wang}},
  \bibinfo {author} {\bibfnamefont {S.}~\bibnamefont {Souma}}, \ and\ \bibinfo
  {author} {\bibnamefont {et~al.}},\ }\bibfield  {title} {\enquote {\bibinfo
  {title} {{Observation of Chiral Fermions with a Large Topological Charge and
  Associated Fermi-Arc Surface States in CoSi}},}\ }\href {\doibase
  10.1103/PhysRevLett.122.076402} {\bibfield  {journal} {\bibinfo  {journal}
  {Phys. Rev. Lett.}\ }\textbf {\bibinfo {volume} {122}},\ \bibinfo {pages}
  {076402} (\bibinfo {year} {2019})}\BibitemShut {NoStop}%
\bibitem [{\citenamefont {Rao}\ \emph {et~al.}(2019)\citenamefont {Rao},
  \citenamefont {Li}, \citenamefont {Zhang},\ and\ \citenamefont
  {et~al.}}]{rao2019observation}%
  \BibitemOpen
  \bibfield  {author} {\bibinfo {author} {\bibfnamefont {Z.}~\bibnamefont
  {Rao}}, \bibinfo {author} {\bibfnamefont {H.}~\bibnamefont {Li}}, \bibinfo
  {author} {\bibfnamefont {T.}~\bibnamefont {Zhang}}, \ and\ \bibinfo {author}
  {\bibnamefont {et~al.}},\ }\bibfield  {title} {\enquote {\bibinfo {title}
  {{Observation of unconventional chiral fermions with long Fermi arcs in
  CoSi}},}\ }\href {\doibase 10.1038/s41586-019-1031-8} {\bibfield  {journal}
  {\bibinfo  {journal} {Nature}\ ,\ \bibinfo {pages} {1}} (\bibinfo {year}
  {2019})}\BibitemShut {NoStop}%
\bibitem [{\citenamefont {Sanchez}\ \emph {et~al.}(2019)\citenamefont
  {Sanchez}, \citenamefont {Belopolski}, \citenamefont {Cochran},\ and\
  \citenamefont {et~al.}}]{sanchez2019topological}%
  \BibitemOpen
  \bibfield  {author} {\bibinfo {author} {\bibfnamefont {D.~S.}\ \bibnamefont
  {Sanchez}}, \bibinfo {author} {\bibfnamefont {I.}~\bibnamefont {Belopolski}},
  \bibinfo {author} {\bibfnamefont {T.~A.}\ \bibnamefont {Cochran}}, \ and\
  \bibinfo {author} {\bibnamefont {et~al.}},\ }\bibfield  {title} {\enquote
  {\bibinfo {title} {{Topological chiral crystals with helicoid-arc quantum
  states}},}\ }\href {\doibase 10.1038/s41586-019-1037-2} {\bibfield  {journal}
  {\bibinfo  {journal} {Nature}\ }\textbf {\bibinfo {volume} {567}},\ \bibinfo
  {pages} {500} (\bibinfo {year} {2019})}\BibitemShut {NoStop}%
\bibitem [{\citenamefont {Schr{\"o}ter}\ \emph {et~al.}(2019)\citenamefont
  {Schr{\"o}ter}, \citenamefont {Pei}, \citenamefont {Vergniory},\ and\
  \citenamefont {et~al.}}]{schroter2019chiral}%
  \BibitemOpen
  \bibfield  {author} {\bibinfo {author} {\bibfnamefont {N.~B.~M.}\
  \bibnamefont {Schr{\"o}ter}}, \bibinfo {author} {\bibfnamefont {Ding}\
  \bibnamefont {Pei}}, \bibinfo {author} {\bibfnamefont {M.~G.}\ \bibnamefont
  {Vergniory}}, \ and\ \bibinfo {author} {\bibnamefont {et~al.}},\ }\bibfield
  {title} {\enquote {\bibinfo {title} {{Chiral topological semimetal with
  multifold band crossings and long Fermi arcs}},}\ }\href {\doibase
  10.1038/s41567-019-0511-y} {\bibfield  {journal} {\bibinfo  {journal} {Nature
  Phys.}\ }\textbf {\bibinfo {volume} {567}},\ \bibinfo {pages} {496} (\bibinfo
  {year} {2019})}\BibitemShut {NoStop}%
\bibitem [{\citenamefont {Chang}\ \emph {et~al.}(2017)\citenamefont {Chang},
  \citenamefont {Xu}, \citenamefont {Wieder},\ and\ \citenamefont
  {et~al.}}]{PhysRevLett.119.206401}%
  \BibitemOpen
  \bibfield  {author} {\bibinfo {author} {\bibfnamefont {G.}~\bibnamefont
  {Chang}}, \bibinfo {author} {\bibfnamefont {S.-Y.}\ \bibnamefont {Xu}},
  \bibinfo {author} {\bibfnamefont {B.~J.}\ \bibnamefont {Wieder}}, \ and\
  \bibinfo {author} {\bibnamefont {et~al.}},\ }\bibfield  {title} {\enquote
  {\bibinfo {title} {{Unconventional Chiral Fermions and Large Topological
  Fermi Arcs in RhSi}},}\ }\href {\doibase 10.1103/PhysRevLett.119.206401}
  {\bibfield  {journal} {\bibinfo  {journal} {Phys. Rev. Lett.}\ }\textbf
  {\bibinfo {volume} {119}},\ \bibinfo {pages} {206401} (\bibinfo {year}
  {2017})}\BibitemShut {NoStop}%
\bibitem [{\citenamefont {Flicker}\ \emph {et~al.}(2018)\citenamefont
  {Flicker}, \citenamefont {de~Juan}, \citenamefont {Bradlyn},\ and\
  \citenamefont {et~al.}}]{PhysRevB.98.155145}%
  \BibitemOpen
  \bibfield  {author} {\bibinfo {author} {\bibfnamefont {F.}~\bibnamefont
  {Flicker}}, \bibinfo {author} {\bibfnamefont {F.}~\bibnamefont {de~Juan}},
  \bibinfo {author} {\bibfnamefont {B.}~\bibnamefont {Bradlyn}}, \ and\
  \bibinfo {author} {\bibnamefont {et~al.}},\ }\bibfield  {title} {\enquote
  {\bibinfo {title} {{Chiral optical response of multifold fermions}},}\ }\href
  {\doibase 10.1103/PhysRevB.98.155145} {\bibfield  {journal} {\bibinfo
  {journal} {Phys. Rev. B}\ }\textbf {\bibinfo {volume} {98}},\ \bibinfo
  {pages} {155145} (\bibinfo {year} {2018})}\BibitemShut {NoStop}%
\bibitem [{\citenamefont {S\'anchez-Mart\'{\i}nez}\ \emph
  {et~al.}(2019)\citenamefont {S\'anchez-Mart\'{\i}nez}, \citenamefont
  {de~Juan},\ and\ \citenamefont {Grushin}}]{PhysRevB.99.155145}%
  \BibitemOpen
  \bibfield  {author} {\bibinfo {author} {\bibfnamefont {M.-A.}\ \bibnamefont
  {S\'anchez-Mart\'{\i}nez}}, \bibinfo {author} {\bibfnamefont
  {F.}~\bibnamefont {de~Juan}}, \ and\ \bibinfo {author} {\bibfnamefont
  {A.~G.}\ \bibnamefont {Grushin}},\ }\bibfield  {title} {\enquote {\bibinfo
  {title} {{Linear optical conductivity of chiral multifold fermions}},}\
  }\href {\doibase 10.1103/PhysRevB.99.155145} {\bibfield  {journal} {\bibinfo
  {journal} {Phys. Rev. B}\ }\textbf {\bibinfo {volume} {99}},\ \bibinfo
  {pages} {155145} (\bibinfo {year} {2019})}\BibitemShut {NoStop}%
\bibitem [{\citenamefont {D\'ora}\ \emph {et~al.}(2011)\citenamefont {D\'ora},
  \citenamefont {Kailasvuori},\ and\ \citenamefont
  {Moessner}}]{PhysRevB.84.195422}%
  \BibitemOpen
  \bibfield  {author} {\bibinfo {author} {\bibfnamefont {B.}~\bibnamefont
  {D\'ora}}, \bibinfo {author} {\bibfnamefont {J.}~\bibnamefont {Kailasvuori}},
  \ and\ \bibinfo {author} {\bibfnamefont {R.}~\bibnamefont {Moessner}},\
  }\bibfield  {title} {\enquote {\bibinfo {title} {Lattice generalization of
  the dirac equation to general spin and the role of the flat band},}\ }\href
  {\doibase 10.1103/PhysRevB.84.195422} {\bibfield  {journal} {\bibinfo
  {journal} {Phys. Rev. B}\ }\textbf {\bibinfo {volume} {84}},\ \bibinfo
  {pages} {195422} (\bibinfo {year} {2011})}\BibitemShut {NoStop}%
\bibitem [{\citenamefont {Raoux}\ \emph {et~al.}(2014)\citenamefont {Raoux},
  \citenamefont {Morigi}, \citenamefont {Fuchs}, \citenamefont {Pi\'echon},\
  and\ \citenamefont {Montambaux}}]{PhysRevLett.112.026402}%
  \BibitemOpen
  \bibfield  {author} {\bibinfo {author} {\bibfnamefont {A.}~\bibnamefont
  {Raoux}}, \bibinfo {author} {\bibfnamefont {M.}~\bibnamefont {Morigi}},
  \bibinfo {author} {\bibfnamefont {J.-N.}\ \bibnamefont {Fuchs}}, \bibinfo
  {author} {\bibfnamefont {F.}~\bibnamefont {Pi\'echon}}, \ and\ \bibinfo
  {author} {\bibfnamefont {G.}~\bibnamefont {Montambaux}},\ }\bibfield  {title}
  {\enquote {\bibinfo {title} {{From Dia- to Paramagnetic Orbital
  Susceptibility of Massless Fermions}},}\ }\href {\doibase
  10.1103/PhysRevLett.112.026402} {\bibfield  {journal} {\bibinfo  {journal}
  {Phys. Rev. Lett.}\ }\textbf {\bibinfo {volume} {112}},\ \bibinfo {pages}
  {026402} (\bibinfo {year} {2014})}\BibitemShut {NoStop}%
\bibitem [{\citenamefont {Vigh}\ \emph {et~al.}(2013)\citenamefont {Vigh},
  \citenamefont {Oroszl\'any}, \citenamefont {Vajna}, \citenamefont {San-Jose},
  \citenamefont {D\'avid}, \citenamefont {Cserti},\ and\ \citenamefont
  {D\'ora}}]{PhysRevB.88.161413}%
  \BibitemOpen
  \bibfield  {author} {\bibinfo {author} {\bibfnamefont {M.}~\bibnamefont
  {Vigh}}, \bibinfo {author} {\bibfnamefont {L.}~\bibnamefont {Oroszl\'any}},
  \bibinfo {author} {\bibfnamefont {S.}~\bibnamefont {Vajna}}, \bibinfo
  {author} {\bibfnamefont {P.}~\bibnamefont {San-Jose}}, \bibinfo {author}
  {\bibfnamefont {G.}~\bibnamefont {D\'avid}}, \bibinfo {author} {\bibfnamefont
  {J.}~\bibnamefont {Cserti}}, \ and\ \bibinfo {author} {\bibfnamefont
  {B.}~\bibnamefont {D\'ora}},\ }\bibfield  {title} {\enquote {\bibinfo {title}
  {{Diverging dc conductivity due to a flat band in a disordered system of
  pseudospin-1 Dirac-Weyl fermions}},}\ }\href {\doibase
  10.1103/PhysRevB.88.161413} {\bibfield  {journal} {\bibinfo  {journal} {Phys.
  Rev. B}\ }\textbf {\bibinfo {volume} {88}},\ \bibinfo {pages} {161413}
  (\bibinfo {year} {2013})}\BibitemShut {NoStop}%
\bibitem [{\citenamefont {Malcolm}\ and\ \citenamefont
  {Nicol}(2015)}]{PhysRevB.92.035118}%
  \BibitemOpen
  \bibfield  {author} {\bibinfo {author} {\bibfnamefont {J.~D.}\ \bibnamefont
  {Malcolm}}\ and\ \bibinfo {author} {\bibfnamefont {E.~J.}\ \bibnamefont
  {Nicol}},\ }\bibfield  {title} {\enquote {\bibinfo {title} {{Magneto-optics
  of massless Kane fermions: Role of the flat band and unusual Berry phase}},}\
  }\href {\doibase 10.1103/PhysRevB.92.035118} {\bibfield  {journal} {\bibinfo
  {journal} {Phys. Rev. B}\ }\textbf {\bibinfo {volume} {92}},\ \bibinfo
  {pages} {035118} (\bibinfo {year} {2015})}\BibitemShut {NoStop}%
\bibitem [{\citenamefont {Biswas}\ and\ \citenamefont
  {Ghosh}(2016)}]{Biswas_2016}%
  \BibitemOpen
  \bibfield  {author} {\bibinfo {author} {\bibfnamefont {T.}~\bibnamefont
  {Biswas}}\ and\ \bibinfo {author} {\bibfnamefont {T.~K.}\ \bibnamefont
  {Ghosh}},\ }\bibfield  {title} {\enquote {\bibinfo {title} {{Magnetotransport
  properties of the$\alpha$-T3 model}},}\ }\href {\doibase
  10.1088/0953-8984/28/49/495302} {\bibfield  {journal} {\bibinfo  {journal}
  {Journal of Physics: Cond. Mat.}\ }\textbf {\bibinfo {volume} {28}},\
  \bibinfo {pages} {495302} (\bibinfo {year} {2016})}\BibitemShut {NoStop}%
\bibitem [{\citenamefont {Dey}\ and\ \citenamefont
  {Ghosh}(2019)}]{PhysRevB.99.205429}%
  \BibitemOpen
  \bibfield  {author} {\bibinfo {author} {\bibfnamefont {B.}~\bibnamefont
  {Dey}}\ and\ \bibinfo {author} {\bibfnamefont {T.~K.}\ \bibnamefont
  {Ghosh}},\ }\bibfield  {title} {\enquote {\bibinfo {title} {{Floquet
  topological phase transition in the $\alpha-{T}_{3}$ lattice}},}\ }\href
  {\doibase 10.1103/PhysRevB.99.205429} {\bibfield  {journal} {\bibinfo
  {journal} {Phys. Rev. B}\ }\textbf {\bibinfo {volume} {99}},\ \bibinfo
  {pages} {205429} (\bibinfo {year} {2019})}\BibitemShut {NoStop}%
\bibitem [{\citenamefont {Eckardt}(2017)}]{RevModPhys.89.011004}%
  \BibitemOpen
  \bibfield  {author} {\bibinfo {author} {\bibfnamefont {A.}~\bibnamefont
  {Eckardt}},\ }\bibfield  {title} {\enquote {\bibinfo {title} {{Colloquium:
  Atomic quantum gases in periodically driven optical lattices}},}\ }\href
  {\doibase 10.1103/RevModPhys.89.011004} {\bibfield  {journal} {\bibinfo
  {journal} {Rev. Mod. Phys.}\ }\textbf {\bibinfo {volume} {89}},\ \bibinfo
  {pages} {011004} (\bibinfo {year} {2017})}\BibitemShut {NoStop}%
\bibitem [{\citenamefont {Yariv}\ and\ \citenamefont {Yeh}()}]{Yariv}%
  \BibitemOpen
  \bibfield  {author} {\bibinfo {author} {\bibfnamefont {A.}~\bibnamefont
  {Yariv}}\ and\ \bibinfo {author} {\bibfnamefont {P.}~\bibnamefont {Yeh}},\
  }\bibfield  {title} {\enquote {\bibinfo {title} {Optical waves in crystal},}\
  }\href@noop {} {\bibinfo  {journal} {Optical waves in crystal (Wiley, New
  York 1984)}\ }\BibitemShut {NoStop}%
\bibitem [{\citenamefont {Seshadri}(1962)}]{Seshadri}%
  \BibitemOpen
\bibfield  {journal} {  }\bibfield  {author} {\bibinfo {author} {\bibfnamefont
  {S.~R.}\ \bibnamefont {Seshadri}},\ }\bibfield  {title} {\enquote {\bibinfo
  {title} {{Excitation of surface waves on a perfectly conducting screen
  covered with anisotropic plasma}},}\ }\href {\doibase
  10.1109/IRETMTT.1962.7527114} {\bibfield  {journal} {\bibinfo  {journal} {IRE
  Trans. Microwave Theory. Tech.}\ }\textbf {\bibinfo {volume} {10}},\ \bibinfo
  {pages} {573} (\bibinfo {year} {1962})}\BibitemShut {NoStop}%
\bibitem [{\citenamefont {Zhukov}\ and\ \citenamefont
  {Raikh}(2000)}]{PhysRevB.61.12842}%
  \BibitemOpen
  \bibfield  {author} {\bibinfo {author} {\bibfnamefont {L.~E.}\ \bibnamefont
  {Zhukov}}\ and\ \bibinfo {author} {\bibfnamefont {M.~E.}\ \bibnamefont
  {Raikh}},\ }\bibfield  {title} {\enquote {\bibinfo {title} {{Chiral
  electromagnetic waves at the boundary of optical isomers: Quantum
  Cotton-Mouton effect}},}\ }\href {\doibase 10.1103/PhysRevB.92.115310}
  {\bibfield  {journal} {\bibinfo  {journal} {Phys. Rev. B}\ }\textbf {\bibinfo
  {volume} {61}},\ \bibinfo {pages} {12842--12847} (\bibinfo {year}
  {2000})}\BibitemShut {NoStop}%
\bibitem [{\citenamefont {Zyuzin}\ and\ \citenamefont
  {Zyuzin}(2015)}]{PhysRevB.92.115310}%
  \BibitemOpen
  \bibfield  {author} {\bibinfo {author} {\bibfnamefont {A.~A.}\ \bibnamefont
  {Zyuzin}}\ and\ \bibinfo {author} {\bibfnamefont {V.~A.}\ \bibnamefont
  {Zyuzin}},\ }\bibfield  {title} {\enquote {\bibinfo {title} {{Chiral
  electromagnetic waves in Weyl semimetals}},}\ }\href {\doibase
  10.1103/PhysRevB.91.241404} {\bibfield  {journal} {\bibinfo  {journal} {Phys.
  Rev. B}\ }\textbf {\bibinfo {volume} {92}},\ \bibinfo {pages} {115310}
  (\bibinfo {year} {2015})}\BibitemShut {NoStop}%
\bibitem [{\citenamefont {Landau}\ and\ \citenamefont {Lifshitz}()}]{Landau}%
  \BibitemOpen
  \bibfield  {author} {\bibinfo {author} {\bibfnamefont {L.~D.}\ \bibnamefont
  {Landau}}\ and\ \bibinfo {author} {\bibfnamefont {E.~M.}\ \bibnamefont
  {Lifshitz}},\ }\bibfield  {title} {\enquote {\bibinfo {title} {{Quantum
  Mechanics}},}\ }\href@noop {} {\bibinfo  {journal} {Pergamon, New York
  1977)}\ }\BibitemShut {NoStop}%
\bibitem [{\citenamefont {Calvo}\ \emph {et~al.}(2015)\citenamefont {Calvo},
  \citenamefont {Foa~Torres}, \citenamefont {Perez-Piskunow}, \citenamefont
  {Balseiro},\ and\ \citenamefont {Usaj}}]{PhysRevB.91.241404}%
  \BibitemOpen
\bibfield  {journal} {  }\bibfield  {author} {\bibinfo {author} {\bibfnamefont
  {H.~L.}\ \bibnamefont {Calvo}}, \bibinfo {author} {\bibfnamefont {L.~E.~F.}\
  \bibnamefont {Foa~Torres}}, \bibinfo {author} {\bibfnamefont {P.~M.}\
  \bibnamefont {Perez-Piskunow}}, \bibinfo {author} {\bibfnamefont {C.~A.}\
  \bibnamefont {Balseiro}}, \ and\ \bibinfo {author} {\bibfnamefont
  {G.}~\bibnamefont {Usaj}},\ }\bibfield  {title} {\enquote {\bibinfo {title}
  {{Floquet interface states in illuminated three-dimensional topological
  insulators}},}\ }\href {\doibase 10.1103/PhysRevB.91.241404} {\bibfield
  {journal} {\bibinfo  {journal} {Phys. Rev. B}\ }\textbf {\bibinfo {volume}
  {91}},\ \bibinfo {pages} {241404} (\bibinfo {year} {2015})}\BibitemShut
  {NoStop}%
\bibitem [{\citenamefont {Menon}\ \emph {et~al.}(2018)\citenamefont {Menon},
  \citenamefont {Chowdhury},\ and\ \citenamefont {Basu}}]{PhysRevB.98.205109}%
  \BibitemOpen
  \bibfield  {author} {\bibinfo {author} {\bibfnamefont {A.}~\bibnamefont
  {Menon}}, \bibinfo {author} {\bibfnamefont {D.}~\bibnamefont {Chowdhury}}, \
  and\ \bibinfo {author} {\bibfnamefont {B.}~\bibnamefont {Basu}},\ }\bibfield
  {title} {\enquote {\bibinfo {title} {Photoinduced tunable anomalous hall and
  nernst effects in tilted weyl semimetals using floquet theory},}\ }\href
  {\doibase 10.1103/PhysRevB.98.205109} {\bibfield  {journal} {\bibinfo
  {journal} {Phys. Rev. B}\ }\textbf {\bibinfo {volume} {98}},\ \bibinfo
  {pages} {205109} (\bibinfo {year} {2018})}\BibitemShut {NoStop}%
\end{thebibliography}%
\end{document}